\documentclass[prd,twocolumn,10pt,
superscriptaddress,
preprintnumbers,
nofootinbib,
]{revtex4-2}

\usepackage{aas_macros}
\usepackage{graphicx}
\usepackage{dcolumn}
\usepackage{bm}
\usepackage[utf8]{inputenc}
\usepackage{amsmath,amssymb}
\usepackage{graphicx}
\usepackage[dvipsnames]{xcolor}
\usepackage{hyperref}
\usepackage{soul}
\usepackage{verbatim}
\usepackage{multirow}
\usepackage{orcidlink}
\usepackage{comment}
\usepackage[normalem]{ulem}

\hypersetup{
  colorlinks   = true, 
  urlcolor     = ForestGreen, 
  linkcolor    = PineGreen, 
  citecolor   = ForestGreen 
}

\begin{document}

\preprint{PITT-PACC-2511}

\title{Galactic magnetic fields seeded by ultralight dark photons}

\author{Joshua Berger\orcidlink{0000-0003-0276-1770}}
\email{joshua.berger@colostate.edu}
\affiliation{%
 Department of Physics, Colorado State University, Fort Collins, CO 80523, United States\\
 }%
 
\author{Amit Bhoonah\orcidlink{0000-0002-4206-215X}}
\email{amit.bhoonah@pitt.edu}
\affiliation{%
 Pittsburgh Particle Physics, Astrophysics, and Cosmology Center, Department of Physics and Astronomy, University of Pittsburgh, Pittsburgh, PA 15260, United States\\
 }%
 
\author{Joseph Bramante\orcidlink{0000-0001-8905-1960}}
\email{joseph.bramante@queensu.ca}
\affiliation{Department of Physics, Engineering Physics, and Astronomy, Queen’s University, Kingston, ON K7L 3N6, Canada}
\affiliation{Arthur B. McDonald Canadian Astroparticle Physics Research Institute, Kingston, ON K7L 3N6, Canada}
\affiliation{Perimeter Institute for Theoretical Physics, Waterloo, ON N2L 2Y5, Canada}

\author{\\J. Leo Kim\orcidlink{0000-0001-8699-834X}}
\email{jlkim@yorku.ca}
\affiliation{Department of Physics and Astronomy, York University, Toronto, ON M3J 1P3, Canada}
\affiliation{Department of Physics, Engineering Physics, and Astronomy, Queen’s University, Kingston, ON K7L 3N6, Canada}
\affiliation{Arthur B. McDonald Canadian Astroparticle Physics Research Institute, Kingston, ON K7L 3N6, Canada}

\author{Ningqiang Song\orcidlink{0000-0002-3590-2341}}%
\email{songnq@itp.ac.cn}
\affiliation{Institute of Theoretical Physics, Chinese Academy of Sciences, Beijing, 100190, China
}%

\author{Lawrence M. Widrow\orcidlink{0000-0001-6211-8635}}%
\email{widrow@queensu.ca}
\affiliation{Department of Physics, Engineering Physics, and Astronomy, Queen’s University, Kingston, ON K7L 3N6, Canada}

\date{\today}

\begin{abstract}
In this work, we show that ultralight dark photons, which couple to the Standard Model photon through kinetic mixing, can potentially source galactic scale magnetic fields. Although these magnetic fields would be too weak to detect at present in galaxies due to plasma screening effects, we show that dark photons can provide the seed magnetic field strength ($10^{-20}$ G) required for dynamo amplification in galaxies. Such dynamo-amplified magnetic fields are consistent with observations of $\mu$G strength galactic magnetic fields. 

\end{abstract}

\maketitle

\section{\label{sec:level1}Introduction}

Magnetic fields are present at many levels of structure in the Universe. They have been observed in low redshift galaxies with strengths of $\mathcal{O}(10) \ \mu \text{G}$ over galactic scales of $\mathcal{O}$(kpc) \cite{Beck2009ASTRA, Beck_microgauss}. On larger scales, $ \mathcal{O}(1)$ $\mu$G fields with $\mathcal{O}$(Mpc) coherence lengths have been observed in galaxy clusters \cite{Carilli:2001hj, Osinga:2022tos,Hu:2023hve}. While the exact origin of magnetic fields over galactic length scales is not fully understood, a commonly accepted scenario is that they are the result of a cosmological seed field which gets exponentially amplified by the action of a dynamo operating during structure formation \cite{2002RvMP...74..775W,Brandenburg2023}. Even though no direct evidence exists for such extragalactic scale magnetic seed fields, this possibility currently cannot be excluded. The seed field hypothesis can accommodate varying strengths of magnetic fields over different length scales, since within a galaxy the seed is assumed to be strongly amplified compared to intracluster regions where the dynamo action is less efficient. There is a vast body of literature on the possible cosmological origin of large-scale seed fields, such as from inflation \cite{Turner:1987bw, Ratra:1991bn}, phase transitions \cite{Kahniashvili:2012uj}, axion electrodynamics \cite{Brandenberger:2025gks}, and magnetic field transfer from a hidden sector via kinetic mixing \cite{Kamada:2018kyi} -- for a review see $e.g.$ Ref.~\cite{Subramanian:2015lua}. These various realizations of the large-scale seed field hypothesis provide an economical cosmological explanation for the observation of magnetic fields across intergalactic, intracluster, and intercluster distances. However, since the existence of magnetic fields on \emph{all} these length scales is itself still an open question, it is interesting to consider the possibility that magnetic fields of different coherence lengths have different origins.

This work investigates how very light, $i.e.$ \emph{fuzzy}, bosonic dark matter (DM) fields could contribute to the generation of galactic magnetic fields. Fuzzy dark matter (FDM) was originally proposed as a resolution to certain shortcomings found when comparing simulations of Cold Dark Matter (CDM) \cite{Hu:2000ke,Hui:2016ltb} with observations, specifically dwarf galaxies with cores \cite{1994Natur.370..629M} and missing satellite galaxies \cite{1993MNRAS.264..201K,1999ApJ...522...82K,2011MNRAS.415L..40B}. The current understanding of these issues is that they are solved by improved observations \cite{Kim:2017iwr} and including baryonic effects in galaxy formation simulations \cite{Chan:2015tna}. Nevertheless, FDM presents an interesting paradigm where DM could affect the large-scale structure of galaxies (see Refs.~\cite{Ferreira:2020fam, Hui:2021tkt} for a review). Although FDM is typically modeled by an ultralight scalar field, which we define as having a mass of $\mathcal{O}(10^{-15})$ eV or lighter, vector FDM models have also been studied \cite{Nelson:2011sf, Adshead:2021kvl}. These models also result in novel large-scale gravitational phenomena that can be distinguishable from scalar FDM, such as inducing deterministic timing delays in pulse arrival times detectable through Pulsar Timing Arrays (PTAs)  \cite{Nomura:2019cvc, PPTA:2022eul, Unal:2022ooa, Omiya:2023bio,NANOGrav:2023hvm, Nomura:2025jur} and astrometry \cite{An:2024axz}. They also act as a long-range fifth force that can potentially modify the dynamics of binary systems \cite{LopezNacir:2018epg}.  \\

By introducing non-gravitational couplings through a kinetic mixing between the dark photon and the Standard Model (SM) photon \cite{Holdom:1985ag}, the dark photon can source observable magnetic fields, particularly on galactic and extragalactic scales \cite{Arias:2012az, Dubovsky:2015cca, Kovetz:2018zes}. Upon halo virialization in the early Universe, these dark photons can act as seed fields required for dynamo amplification. We will soon show that these ULDPs with masses of $ \lesssim \mathcal{O}(10^{-21})$ eV can naturally produce magnetic fields coherent over galactic scales. The fields are sourced in a similar way to well-known mechanisms for astrophysical galactic seed fields, which can be produced through the separation of ions and electrons in proto-galactic plasma --- an effect known as the Biermann battery \cite{Biermann1950}. This mechanism is distinct from the cosmological seed field hypothesis, which posits a seed field coherent over length scales larger than Mpcs that are generated well before structure formation.  

In our scenario, magnetic fields are generated only after an over-dense region of the universe virializes to form a proto-galaxy. The overall strength of the ULDP seed field is set by fraction of DM energy density of the proto-galaxy in the form of ULDPs\footnote{For reasons that will be explained later, we will generally assume that DM is an admixture of CDM and ULDPs, the latter not comprising more than 1$\%$ of all the DM in the universe.}, its free electron content, and the coupling strength between ULDPs and normal matter. Interestingly, one can expect more variation in the magnetic field strengths across different galaxies compared to those that would be generated due to cosmological seed fields as, in our case, stochastic variations in the initial abundance of DM and the free electron fraction in the over-density in question predict different magnetic field strengths in different galaxies. Another intriguing consequence of our mechanism is that dwarf galaxies, which have lower baryonic content compared to their normally sized counterparts, should have comparatively weaker fields. \\

The remainder of this work is organized as follows. In Section \ref{sec:uldps} we introduce ULDPs and show how their coupling to SM photons can give rise to electric and magnetic fields. In Section \ref{sec:seed_fields} we consider magnetic fields from ULDPs during the formation of galaxies, in which they can provide the seed field which would eventually be amplified through the dynamo mechanism. In Section \ref{sec:conclusion} we make concluding remarks and discuss limitations and future considerations. Throughout this paper, we use natural units where $\hbar = c = k_B=1$.

\section{Dark electromagnetic fields from ultralight dark photons} \label{sec:uldps}

The model we consider requires a simple extension of the SM by a U(1) symmetry, which means that it has another abelian gauge boson, $A^{\prime \mu}$ --- the ULDP --- in addition to the gauge bosons in the SM. For dynamical processes occurring in the sub-GeV range, the physical $A^{\prime \mu}$ field mixes predominantly with the SM photon $A^\mu$, and the resulting Lagrangian can be written as
\begin{equation}\label{eq:MainLagrangian}
\begin{split}
    \mathcal{L} =  &-\frac{1}{4}F_{\mu \nu} F^{\mu\nu} -\frac{1}{4}F'_{\mu \nu} F'^{\mu\nu} + m^{2}_{\gamma^{\prime}}A_\mu^{\prime} A^{\prime \mu}  \\
    &  -  e\left(A_{\mu} + \epsilon A^{\prime}_{\mu}\right)J^{\mu}.
\end{split}
\end{equation}  
Here $\epsilon$ is the kinetic mixing parameter, $m_{\gamma^{\prime}}$ the mass of the dark photon, and $F_{\mu \nu} = \partial_\mu A_\nu - \partial_\nu A_\mu$ is the usual field strength tensor for the photon with $F_{\mu\nu}'$ similarly defined for the dark photon. We work in the canonical mass basis, where the mixing with the $Z$ boson is negligible for ULDPs. The interaction part of this Lagrangian consists of the photon and the dark photon, both coupled to the electromagnetic 4-current $J^{\mu}$, with the latter coupling suppressed by a factor of $\epsilon$. We comment that although other couplings between dark photons and the SM are possible (for example through $B$ or $B-L$ couplings), we only consider the kinetic mixing in this work. \\

Many possibilities have been studied for the cosmological production of ULDPs that ultimately comprise the DM of the universe \cite{Graham:2015rva, Dror:2018pdh,Long:2019lwl, Kolb:2020fwh, Nelson:2011sf, Arias:2012az}. Since we are interested in late-time effects, the particular realization in question is not of significant importance to us. However for concreteness, we will focus on the well-studied misalignment mechanism for vector DM \cite{Nelson:2011sf,Arias:2012az}, which we now give a brief account of, providing some comments on its difficulties for producing 100$\%$ of the DM for the mass range we are interested in \cite{Graham:2015rva,Arias:2012az} towards the end of this section. \\ 

Assuming the Universe is described on large scales by a spatially flat Friedmann-Lema\^itre-Robertson-Walker (FLRW) metric,
\begin{equation}
    ds^{2} = dt^{2} - a^{2}\left(t\right)\mathbf{d}\mathbf{x}^{2},
\end{equation}
where $a(t)$ is the scale factor of the Universe, the equation of motion for a non-relativistic dark photon field is given by 
\begin{equation}\label{eq:NonrelDPEqu}
    \frac{d^{2}\mathbf{A}^{\prime}}{dt^{2}} + 3H\frac{d\mathbf{A}^{\prime}}{dt} + m_{\gamma'}^2 \mathbf{A}^{\prime}  = 0,
\end{equation}
where $H \equiv \dot{a}/a$ is the Hubble parameter and we have neglected the dark photon momentum and any back-reaction of charged baryonic matter on the dark photon field. The latter is justified for the dark photon mass range and small value of the kinetic mixing parameter we consider in this work. In particular, for dark photons masses below $10^{-17}$ eV, the in-medium conversion of dark photons to photons that can deplete the abundance of dark photons is negligible since the plasma frequency of the medium never resonantly matches the dark photon mass. In the early universe, $H\left(t\right) \gg m_{\gamma^{\prime}}$ and the field evolution in Eq.~\eqref{eq:NonrelDPEqu} is dominated by the Hubble friction term. The field is over-damped and frozen, such that $d\mathbf{A}^{\prime}/dt \approx 0$. In this case, the ULDP field does not oscillate: it remains at a fixed value until it can enter the horizon, which occurs when the Hubble term becomes comparable to the dark photon mass, $H(t) \simeq m_{\gamma^{\prime}}$. Afterwards, the dark photon field begins to oscillate and the different modes of this oscillation represent a population of non-relativistic massive vector bosons that constitute (at least partially) the DM. For a benchmark mass of 10$^{-21}$ eV, the transition from a over-damped and frozen field to an under-damped and oscillatory once occurs at a redshift $z \sim \mathcal{O}(10^{6})$, or when the temperature of photons was about 200 eV. This is still during the radiation-dominated era, albeit well after Big Bang Nucleosynthesis (BBN) and when electrons have become non-relativistic. \\
 
Once $H < m_{\gamma^{\prime}}$, since the Hubble friction term is a decreasing function of time, it becomes subdominant and Eq.~\eqref{eq:NonrelDPEqu} can be solved using the WKB approximation under the initial condition $d\mathbf{A}^{\prime}/dt \vert_{t=t_{m}} = 0$ to give
\begin{equation}\label{eq:AFieldGeneralSolution}
    \mathbf{A}^{\prime}\left(t\right) = \left(\frac{a_m}{a}\right)^{3/2}\mathbf{A}_{m}^{\prime}\cos\left(m_{\gamma^{\prime}}\left(t-t_{m}\right)\right),
\end{equation}
where $t_{m}$ is the time at which the misalignment condition is fulfilled, $a_m = a(t_m)$. The coherence length $\ell_c$ of this field is given by 
\begin{equation}\label{eq:lcoherence}
    \ell_{c} = \frac{2\pi}{m_{\gamma^{\prime}}v},
\end{equation} 
where $v \equiv \vert \mathbf{v} \vert$ is the amplitude of the velocity vector. The corresponding electric and magnetic fields are defined in the usual way, 
\begin{equation}
\begin{split}
\mathbf{E}^{\prime} & \equiv - \nabla A'^{0} \, - \dot{\mathbf{A}}^{\prime} \\  
\mathbf{B}^{\prime} & \equiv \nabla\times\mathbf{A}^{\prime}.
\end{split}
\end{equation}
The amplitude of the field is set by the requirement that the energy density of the ULDPs match the corresponding fraction $f$ of the DM energy density $\rho_{\rm DM}$. At the misalignment time and over a length scale $\ell_{c}$, 
\begin{equation}\label{eq:A0}
    \frac{1}{2}m_{\gamma^{\prime}}^{2}\vert \mathbf{A}^{\prime}_{m} \vert^{2}\left(1+z_{m}\right)^{-3} \simeq f \rho_\mathrm{DM}(z),
\end{equation}
where $z_{m}$ is the value of the redshift at misalignment (we have replaced the scale factor with redshift using $a(t) = 1/(1+z)$). The electric field over a coherence length is then given by 
\begin{equation}\label{eq:darkEField}
    \mathbf{E}^{\prime} = \mathbf{E_{0}}(z)\sin\left(m_{\gamma^{\prime}}(t - t_{m})\right) \ , 
\end{equation}
with 
\begin{align}
   \vert \mathbf{E_{0}}(z)\vert \equiv  E_{0}(z) = \sqrt{2f\rho_\mathrm{DM}(z)}.
\end{align}
The magnetic field, which is not present at zeroth order in the DM velocity, is given by
\begin{equation}\label{eq:DarkBField}
    \mathbf{B}^{\prime} =   \mathbf{v} E_{0}(z) \cos\left(m_{\gamma^{\prime}} (t - t_m) \right)   ,
\end{equation}
where $\mathbf{v}$ is the velocity of the dark photon field. To obtain this expression, we note the fact that taking the curl of the $\mathbf{A}$ introduces a characteristic length scale $\sim 1/\ell_c$. \\

We will assume throughout this work that the DM consists primarily $\mathcal{O}( 99) \%$ of CDM and that the density of ULDPs closely follows the density of CDM, such that 
\begin{align}\label{eq:fULDP}
    \rho_\mathrm{ULDP} (z) = f \rho_\mathrm{CDM} (z),
\end{align}
with $f\sim 1\%$. Eq.~\eqref{eq:fULDP} should be a valid approximation for such small $f$ as long as the dynamical scales are larger than the coherence length. Although one may expect these ULDPs to have a significantly different picture of structure formation, particularly on small scales \cite{Amin:2022pzv}, we assume that their evolution is predominantly determined by the more-abundant CDM component. In general, considering a smaller fraction of ULDPs is expected to relax constraints which arise from kinematical observations of $e.g.$ dwarf galaxies \cite{Zimmermann:2024xvd}, since the gravitational potential is dominated by the abundant component of DM. \\

Indeed, the behavior of this field over lengths much greater than $\ell_{c}$, or times larger than the coherence time, $\tau_{c} \simeq 2\pi/ m_{\gamma^{\prime}}v^{2}$ is much more complicated. In particular, for virialized dark photons, the velocity distribution induces a phase decoherence at times $t \gg \tau_{C}$, when interference between different modes become relevant and cause the net field to fluctuate stochastically. When considering cosmological structure, one must also take into account the spatial decoherence over length scales larger than $\ell_{c}$. Such a complete study incorporating all these effects is only possible with a proper cosmological simulation, which is beyond the scope of this work. We will therefore assume, for simplicity, that the dark photon has a mass of roughly $\sim10^{-21}$ eV, corresponding to de Broglie wavelength of $\mathcal{O}$ (kpc), and represents a minute fraction of the DM, as previously mentioned. In such a scenario, the dark electric and magnetic fields are well described by Eqs. \eqref{eq:darkEField} and \eqref{eq:DarkBField} over the length and time scales relevant for this work. \\

Before closing this section, we offer some additional remarks regarding our assumption that $f\sim(1\%)$. We assume that the ULDPs only provide a fraction of the observed dark matter density, because this assumption addresses several details: (i) Constraints on the kinetic mixing $\epsilon$ are generally relaxed due to ULDPs not making up all of the DM. For instance, note that Ref.~\cite{Kobayashi:2017jcf} found that Lyman-$\alpha$ constraints on ultralight bosons are lifted for $f \lesssim 0.2$. (ii) As previously mentioned, the dynamics of halo formation and evolution for ULDPs are expected to follow the dynamics of the more-abundant CDM component. (iii) Finally, the difficulties in producing a sufficient abundance of low mass dark photons \cite{Arias:2012az, Graham:2015rva, Agrawal:2018vin} can be alleviated if it is not all of the DM. 

\section{Seed magnetic fields generated by ultralight dark photons} \label{sec:seed_fields}

The mechanism we propose for seeding magnetic fields from ULDPs can be summarized as follows: after the Universe is populated with cold ULDPs that behave like a background dark electromagnetic field, by virtue of the Lorentz force law, charged particles in the cosmic fluid are displaced by the dark electric and magnetic fields. Because of the mobility difference between electrons and ions, a net current is sourced which generates a magnetic field from Amp\`ere's Law. As the dark magnetic field depends on the DM velocity, which is random prior to virialization, no net magnetic field can be generated before that point. The idea is similar to that first proposed by Refs.~\cite{1970MNRAS.147..279H,PhysRevLett.30.188} for generating primeval magnetic fields during the radiation domination era through vorticity. This was later applied to rotating proto-galaxies post-recombination \cite{1972JETP...34..233M}. As we will show, the presence of the dark photon field can generate a seed field comparable to that generated through vorticity effects in a proto-galaxy.

\subsection{Seed dark magnetic fields}

For seed magnetic fields present at times of galaxy formation, we consider the galactic environment near virialization. To estimate the density of ULDPs at times of halo formation, we consider a halo undergoing spherical collapse. For simplicity, we assume that the distribution of ULDPs will trace the distribution of the dominant, cold DM. The density of this halo at redshift $z$ is given by
\begin{align}
    \rho_\mathrm{DM} \sim \Delta_c \rho_{c,0} (1+z)^3,
\end{align}
where $\Delta_c = 18\pi^2 \approx 178$ is the usual overdensity relative to the critical density upon virialization, and $\rho_{c,0} = 3 H_0^2 / (8 \pi G)$ is the present-day background density. We assume Planck values of $\Omega_m = 0.315$ for the matter density parameter, $\Omega_\Lambda = 1 - \Omega_m$ for the density parameter for the cosmological constant $\Lambda$, and $\Omega_b = 0.0221h^{-2} \approx 0.0493$ as the baryon density parameter, where we have assumed $H_0 = 67.4$ km/s/Mpc is the current expansion rate of the Universe \cite{Planck:2018vyg} so that $h = 0.674$ is the dimensionless Hubble constant. The critical density is then given by $\rho_{c,0} = 1.05 \times 10^{-5} h^2$ GeV/cm$^3 \approx 4.77 \times 10^{-6}$ GeV/cm$^3$. Taking the reference formation time of the halo to be $1+z = 20$, the density of the halo can be estimated as
\begin{align}
    \rho_\mathrm{DM} \approx 6.8 \text{ GeV/cm}^3 \left( \frac{1+z}{20} \right)^{3}. \label{eq:rho_h}
\end{align}
For a halo with mass $M$, the virial radius is given by \cite{Barkana:2000fd}
\begin{align}
    r_\mathrm{vir} = 0.51 \text{ kpc} \left( \frac{M}{10^8 M_\odot} \right)^{1/3} \left( \frac{1+z}{20} \right)^{-1} .
\end{align}
We have chosen a reference mass of $M = 10^8 M_\odot$, typical for galaxy-sized DM halos at a redshift of $z= 20$ \cite{Clarke:2003id}. The circular velocity is given by
\begin{align}
    V_c \approx 30 \text{ km s}^{-1}  \left( \frac{M}{10^8 M_\odot} \right)^{1/3} \left( \frac{1+z}{20} \right)^{1/2} , \label{eq:V_c}
\end{align}
while the virial temperature is 
\begin{align}
    T_\mathrm{vir } \approx 6 \times 10^4 \text{ K} \left( \frac{\mu}{1.22} \right) \left( \frac{M}{10^8 M_\odot} \right)^{2/3}  \left( \frac{1+z}{20} \right), \label{eq:T_vir}
\end{align}
where $\mu = 1.22$ for a neutral primordial gas. Using Eqs. \eqref{eq:DarkBField}, \eqref{eq:rho_h}, and \eqref{eq:V_c}, 
the strength of the dark magnetic field is given by
\begin{align}
    B' \approx 6 \times 10^{-6} \text{ G} \left( \frac{f}{0.01} \right)^{1/2}\left( \frac{M}{10^8 M_\odot} \right)^{1/3} \left( \frac{1+z}{20} \right)^{2},
\end{align}
where we have normalized to a $M = 10^8 M_\odot$ halo.

\begin{figure}
\includegraphics[width=\columnwidth]{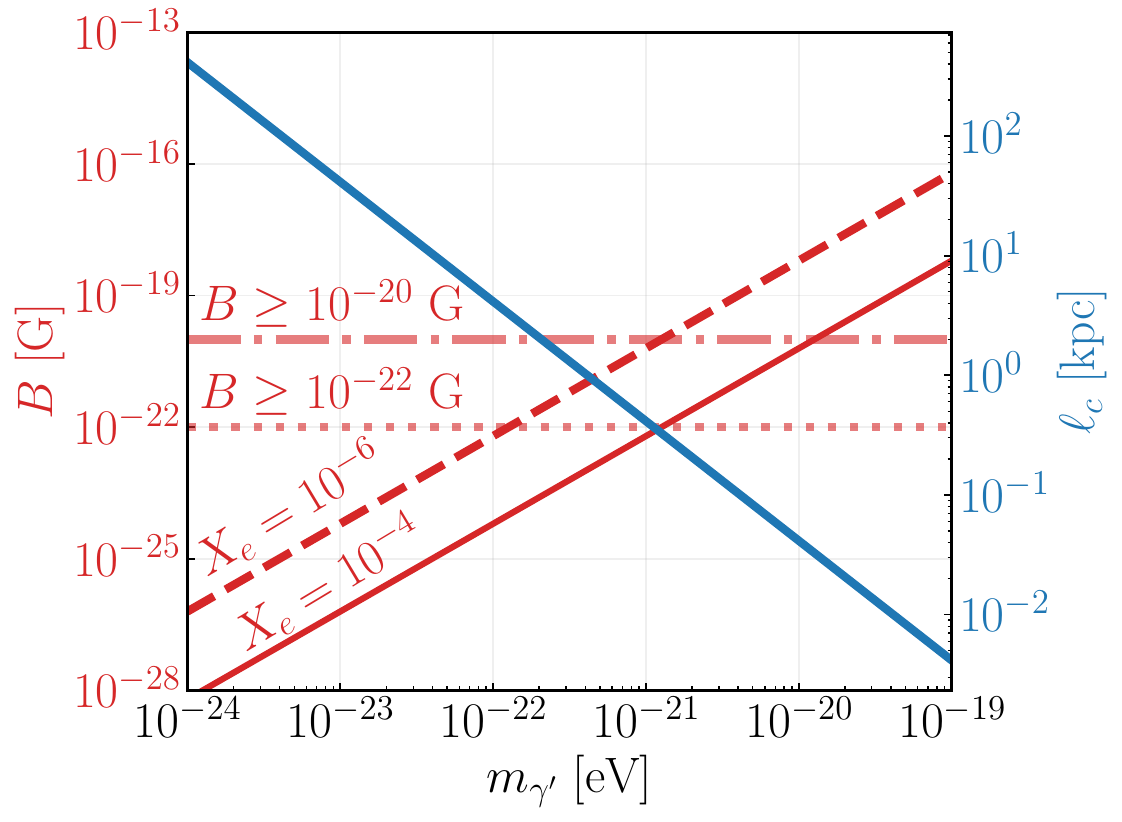}
\caption{\label{fig:B_prim} Magnetic field amplitudes (red) from ULDPs assuming a $f=0.01$ fraction of DM are in ULDPs, with a kinetic mixing of $\epsilon=0.01$. We have assumed two different free electron fractions of $X_e = 10^{-4}$ for the solid line and $X_e = 10^{-6}$ for the dashed line, and taken halo parameters that match a $10^8~M_\odot$ halo at redshift $1+z=20$. Field values above the faded dotted line indicate ideal values which could be amplified via the dynamo effect into the observed galactic magnetic fields. Also plotted are the associated coherence lengths (blue) for the ULDPs. We have indicated rough ``seed field" values of $10^{-20}$ G in dotted-dashed and $10^{-22}$ G in dotted, which are sufficiently large to be amplified to presently observed galactic magnetic field strengths, given the results of simulations in Ref.~\cite{Schober2012}.}
\end{figure}

\subsection{Plasma screening}

To translate this quantity into an observable magnetic field, one needs to taken into account the kinetic mixing between the SM and dark photons. Due to the presence of free electrons inside of the halo however, the observable field does not simply scale with $\epsilon$, but further suppressed due to plasma screening effects \cite{Dubovsky:2015cca}, which depend on the number density of free electrons $n_e$ as well as their temperature $T_e$. We assume that the number density of free electrons in the halo can be estimated by relating the density of baryons to the DM in the Universe at a redshift of $z$, then assuming that we have an approximately neutral gas. Therefore 
\begin{align}
    n_e = X_e \frac{\Omega_b}{\Omega_m} \frac{\rho_{c,0}}{m_p} (1+z)^3 \approx  10^{-6} \text{ cm}^{-3} \left( \frac{X_e}{10^{-4}} \right) \left( \frac{1+z}{20} \right)^3,
\end{align}
where $m_p = 0.938$ GeV is the proton mass and $X_e$ is the free electron abundance. We also assume that the electron temperature is $T_e = T_\mathrm{vir}$, so that the baryonic matter is all thermalized. Note that in this expression, we have taken $n_e$ to depend on the background density of free electrons, rather than the a value matched to the local overdensity of dark matter, as might be assumed instinctively. This is motivated by \cite{Tseliakhovich2010, Naoz2011, Cui2012}, which found that at $z \sim 20$ the residual free electron fraction is affected by the relative baryon-dark matter velocity, and that supersonic baryon–dark matter streaming effects can suppress local enhancements of $n_e$, especially for lower mass halos. 

The density of electrons in the galactic halo is especially relevant, since the observable magnetic field generated by the dark photon is suppressed such that
\begin{align}
    B \approx \epsilon \frac{m_{\gamma'}^2}{\Omega_p^2} B',
\end{align}
where 
\begin{align}
    \Omega_p^2 = \frac{\omega_p^2}{1+ \frac{i\nu}{\omega}},
\end{align}
is the eigenvalue squared of one of the propagating modes in medium \cite{Dubovsky:2015cca}, with $\omega$ the frequency of the wave.  Here $\nu$ is the collisional frequency between electrons and ions, given by
\begin{align}
    \nu &= \frac{4 \sqrt{2\pi} \alpha^2 n_e}{ 3 m_e^{1/2} T_e^{3/2} } \log \Lambda_C \nonumber \\
    &\approx 10^{-27} \text{ eV} \left( \frac{n_e}{10^{-6} \text{ cm$^{-3}$}} \right) \left( \frac{6 \times 10^4 \text{ K}}{T_e} \right)^{3/2},
\end{align}
where $m_e = 511$ keV is the electron mass and $\alpha=1/137$ is the usual fine-structure constant for electromagnetism. The Coulomb logarithm is given by
\begin{align}
    \log \Lambda_C &= \log \left( \frac{4 \pi T_e^3}{\alpha^3 n_e} \right)^{1/2}  \\
    &\approx 35 + \log \left( \frac{T_e}{6 \times 10^4 \text{ K}} \right)^\frac{3}{2} - \log \left( \frac{10^{-6} \text{ cm}^{-3}}{n_e} \right)^\frac{1}{2} . \nonumber
\end{align}
Henceforth, we take $\log \Lambda_c \approx 35$ for simplicity due to the weak dependence on $T_e$ and $n_e$. The plasma frequency
\begin{align}
    \omega_p = \left( \frac{4\pi n_e \alpha}{m_e} \right)^{1/2} \approx 10^{-14} \text{ eV} \left( \frac{n_e}{10^{-6} \text{ cm$^{-3}$}} \right)^{1/2}.
\end{align}
The suppression of the observable magnetic field takes the form 
\begin{align}
    \frac{m_{\gamma'}^2}{\Omega_p^2} \approx
    \begin{cases}
        \frac{m_{\gamma'}^2}{\omega_p^2} &\text{ for } \nu \ll m_{\gamma'} \ll \omega_p, \\
        \frac{i\nu m_{\gamma'}}{\omega_p^2} &\text{ for } m_{\gamma'} \ll \nu \ll \omega_p.
    \end{cases}
\end{align}
This means that for dark photon masses which correspond to de Broglie wavelengths which are $\mathcal{O}$(kpc) scales, we have that we are in the first regime $\nu \ll m_{\gamma'} \ll \omega_p$, and so the temperature dependence of the suppression factor drops out. In this case, following suppression, we have
\begin{align} \label{eq:B_param}
    B \approx &\ 6 \times 10^{-21} \text{ G} \left( \frac{m_{\gamma'}}{10^{-20} \text{ eV}} \right)^{2} \left( \frac{10^{-4}}{X_e} \right)    \\
    &\times  \left( \frac{M}{10^8 M_\odot} \right)^{1/3}  \left( \frac{20}{1+z} \right)^{3/2} \left( \frac{f}{0.01} \right)^{1/2} \left(\frac{\epsilon}{0.01}\right) . \nonumber 
\end{align}
This is the seed field which would be enhanced due to the galactic dynamo.

In principle, one can perform a similar calculation assuming properties of a galaxy and galaxy cluster today, to acquire late-time magnetic fields on scales of $\mathcal{O}$(kpc) and $\mathcal{O}$(Mpc), respectively. These fields from ULDPs can then be compared with observed magnetic field strengths across these scales \cite{Amaral:2021mly}. However, due to the effects of plasma screening, the observable magnetic field strengths from ULDPs are too weak to explain the late-time observed magnetic fields -- implying the necessity of an early dynamo amplification if the magnetic fields from ULDPs were a significant component of the observed magnetic fields.

\subsection{Amplification by the dynamo effect}

In a simplified model, the large-scale dynamo effect amplifies fields such as
\begin{align}
    B \propto e^{\Gamma t},
\end{align}
where $t$ is the duration that the dynamo is active, and $\Gamma$ is the growth rate of magnetic fields. The growth rate should not exceed the orbital period of the galaxy $\Omega$, such that $\Gamma \lesssim \Omega$ \cite{Ruzmaikin_book}. Considering orbital periods anywhere between the Milky Way orbital period of $\Omega \approx 250$ Myr$^{-1}$ to a conservative orbital period of a Gyr$^{-1}$, the enhancement due to the dynamo over a growth time of 10 Gyrs can be anywhere between $e^{10}$ to $e^{40}$. Due to the exponential dependence on the growth rate, the amplification is highly sensitive to the details of small-scale dynamos. Nevertheless, motivated by simulations \cite{Schober2012}, we assume a conservative rough minimum seed field requirement of  $B_\mathrm{min} \approx 10^{-20} ~\text {G}$ as well as an optimistic minimum seed field requirement of $B_\mathrm{min} \approx 10^{-22} ~\text {G}$ to eventually result in the observed late-Universe $\mu$G-scale galactic magnetic fields. We have considered two choices for the seed field requirement because the dynamo enhancement is a highly turbulent process, and so it is heavily dependent on the assumptions entering the simulations. Note that a seed field of this strength can also be generated by astrophysical battery mechanisms, such as the Biermann battery \cite{Biermann1950}.

Fig. \ref{fig:B_prim} shows the observable magnetic field due to ULDPs as a function of their mass. Also plotted are their respective coherence lengths, which were found using Eqs. \eqref{eq:lcoherence} and assuming a virial velocity given by \eqref{eq:V_c}. We have also indicated where a rough minimum field $B_\mathrm{min}$ which would be sufficiently large to be amplified by the dynamo effect to explain the observed galactic $\mu$G-magnetic fields. We have plotted the magnetic field from ULDPs assuming two different free electron fractions of $X_e = 10^{-4}$ (solid) and $X_e = 10^{-6}$ (dashed). As evident in the Figure, the amplitude of the seed field is highly dependent on the fraction of free electrons $X_e$ due to the dependence of the plasma screening on the free electron number density, which suppresses $B$ by a factor of $X_e$ ($c.f.$ Eq. \eqref{eq:B_param}). This is an advantage of considering seed fields prior to reionization which our dynamos start to turn on before, since the free electron abundance is expected to be low before the Universe reionizes.

\begin{figure}
\includegraphics[width=\columnwidth]{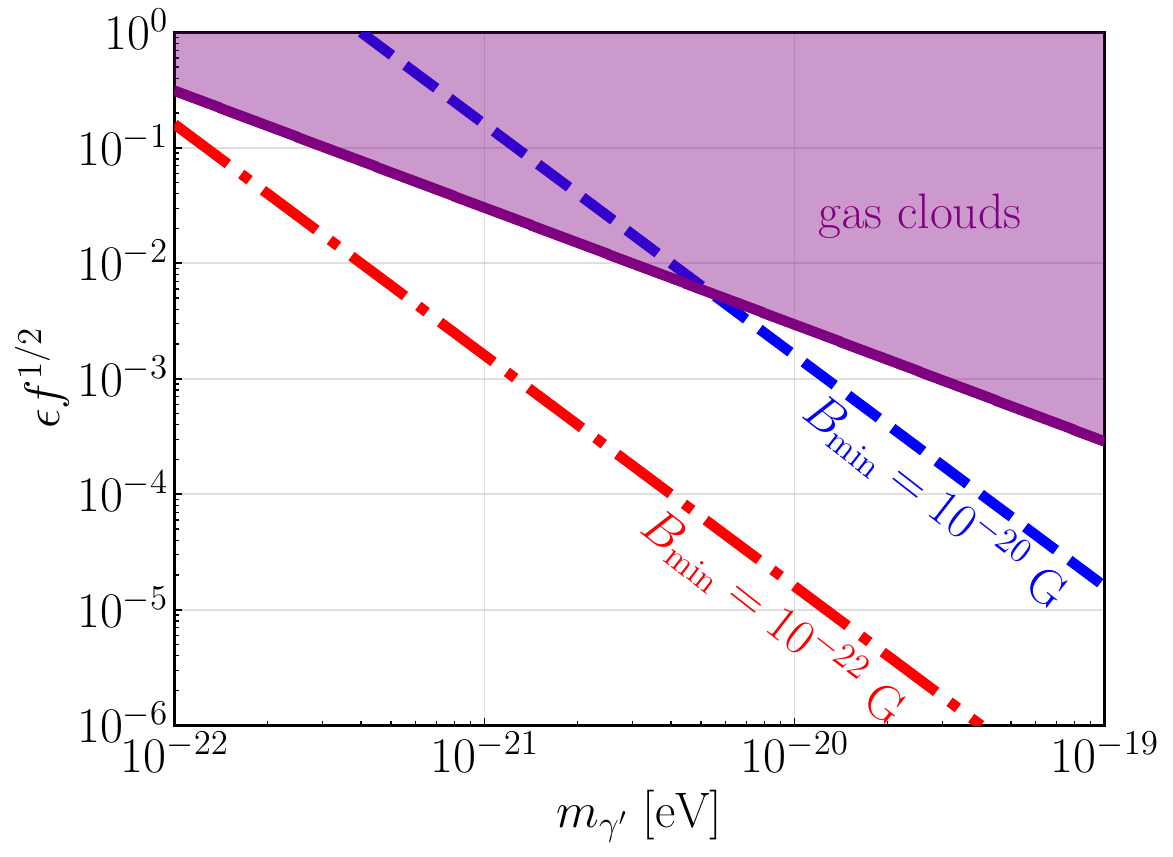}
\caption{\label{fig:B_allowed} Allowed values for the combination of ULDP parameters $\epsilon f^{1/2}$ as a function of the mass $m_{\gamma'}$. The purple shaded area is constrained by the heating of gas clouds \cite{Bhoonah:2018gjb}. The region to the right of the blue dashed (red dotted-dashed) lines indicate parameter space which gives rise to sufficiently strong seed fields, assuming minimum field strengths of $10^{-20}$ G ($10^{-22}$ G). We have assumed a free electron fraction of $X_e = 10^{-4}$ and have taken halo parameters that match a $10^8~M_\odot$ halo at redshift $1+z=20$. }
\end{figure}

Due to the dependence of the dark photon mass on the amount of plasma screening as seen parametrically in Eq. \eqref{eq:B_param}, a higher dark photon mass leads to stronger magnetic fields, although for shorter coherence lengths. Furthermore, as one increases the mass of the ULDP, existing constraints on the kinetic mixing $\epsilon$ start becoming relevant (see $e.g.$, Refs.~\cite{Kovetz:2018zes, Bhoonah:2018gjb, Acevedo:2025ysl}). However, as previously mentioned, since the ULDP in our case is only a subdominant component of the DM, the existing constraints in the mass scales we consider are generally expected to weaken. In Fig. \ref{fig:B_allowed} we show the allowed combination of $f_\mathrm{DM}$ and $\epsilon$ by considering the bounds in Ref.~\cite{Bhoonah:2018gjb} which arise from studying the heating of gas cloud G357.8-4.7-55. The heating rate obtained in Ref.~\cite{Bhoonah:2018gjb} for the heating of gas clouds by ULDPs scales as $Q \sim m_{\gamma'}^2 \epsilon^2 f$, and so we extend their constraints to lower masses. On the other hand, the magnetic field strengths ($c.f.$ Eq. \eqref{eq:B_param}) scales as $B \sim m_{\gamma'}^2 \epsilon f^{1/2}$, leading to a different scalings on ULDP masses. The parameter space to the right of the blue dashed and red dotted-dashed lines in Fig. \ref{fig:B_allowed} show where the magnetic fields from the ULDPs are sufficiently strong to act as seed fields, assuming a minimum field requirement of $B_\mathrm{min} = 10^{-20}$ G and $B_\mathrm{min} = 10^{-22}$ G, respectively. Here we emphasize once more that the minimum field requirement is a rough requirement which is heavily dependent on the assumptions entering the dynamo simulations. At higher values of $f_\mathrm{DM}$ and $\epsilon$, which would lead to stronger magnetic fields, the dark photons become disallowed due to the aforementioned constraints. 

An interesting result is that higher mass ULDPs can produce rather strong seed fields ($\gtrsim10^{-20}$ G) while simultaneously evading constraints from gas cloud heating, due to the difference in parametric dependence on $\epsilon f^{1/2}$. However, as previously discussed, the magnetic field coherence lengths are smaller for higher ULDP masses, with $\ell_c \lesssim 0.1$~kpc for $m_{\gamma'} \gtrsim 10^{-20}$~eV (\textit{cf.}~Fig.~\ref{fig:B_prim}). While the galactic dynamo can amplify these fields, the initial correlation length plays a crucial role. Simple estimates suggest the large-scale mean field constructed from such small-scale seeds is suppressed by a factor of $(\ell_c/L)^{3/2}$, where $L$ is the scale of interest~\cite{Hogan:1983zz, 2002RvMP...74..775W}. However, whether this suppression strictly precludes the generation of observed galactic magnetic fields depends on a more detailed analysis of dynamo efficiency, saturation on these scales, and a more thorough treatment of galaxy mass functions and merger history. We leave such specific dynamo considerations, as well as dependencies on halo formation properties (e.g., DM halo mass and redshift), for future work.

\section{Conclusion} \label{sec:conclusion}

In this work, we have studied how the observed galactic-scale magnetic fields today can potentially arise from a subdominant, ultralight, dark vector boson that kinetically mixes with the Standard Model photon. We demonstrated that ULDPs can source a seed magnetic field while the galaxy virializes, which can eventually be amplified to explain the galactic magnetic fields observed today. Although the complexity of galactic dynamos requires simulations for a complete picture and is therefore an ongoing effort, we have shown that ULDPs can produce seed fields comparable to other astrophysical seed fields, such as one arising from the Biermann battery \cite{Biermann1950}. While we have focused primarily on the possibility that the seed fields from ULDPs are sufficient for dynamo amplification to observed field values, another possible scenario is where the magnetic field from ULDPs is a contributing field in the initial seed fields. 

In calculating the magnetic field from ULDPs, we note that the observable magnetic field amplitude is highly sensitive to the free electron fraction $X_e$ due to plasma screening, and therefore whether a magnetic field from ULDPs can solely be the seed field required for the dynamo effect is also dependent on $X_e$. Although the usual assumption is that the free electron fraction prior to reionization is around $X_e \sim 10^{-4}$, more general model-independent constraints on $X_e$ from the CMB \cite{Villanueva-Domingo:2017ahx, Cheng:2025cmb} typically place upper bounds on $X_e$. While suppressing the abundance of free electrons seems to be difficult in standard cosmological scenarios \cite{Hart:2019gvj}, it is nevertheless an interesting possibility ($e.g.$ \cite{Falkowski:2018qdj}).
The upcoming Square Kilometer Array (SKA) \cite{Koopmans:2015sua} is expected to shed light on the epochs of reionization and cosmic dawn, which will help solidify the constraints on the allowed free electron abundance at high redshifts.

Finally, we comment that in order to truly model the effects of dynamo amplification, simulations will be necessary. Furthermore, the amplification of the seed field can be significantly different in the case of small-scale and large-scale dynamos \cite{Subramanian:2015lua, Subramanian:2019jyd}. While in this work we assumed a simplified amplification due to a large-scale dynamo, it would be interesting to see if the seed fields from ULDPs can also be sufficiently enhanced in other dynamo scenarios. We leave these considerations for future studies.

\begin{acknowledgments}
J.~Berger's work is supported by the U.S.~National Science Foundation under Award No.~2413017. The work of A.~Bhoonah was supported in part by the U.S. Department of Energy under grant No. DE-SC0007914 and in part by the Pittsburgh Particle Physics Astrophysics and Cosmology Center (PITT PACC). J.~Bramante, J.~L.~Kim, and L. M. Widrow were supported by the Natural Sciences and Engineering Research Council of Canada (NSERC) and the Canada Foundation for Innovation. This research was undertaken thanks in part to funding from the Arthur B. McDonald Canadian Astroparticle Physics Research Institute. Research at Perimeter Institute is supported by the Government of Canada through the Department of Innovation, Science, and Economic Development, and by the Province of Ontario. N. Song is supported by the National Natural Science Foundation of China (NSFC) under Grant Nos. 12475110, 12347105, 12441504 and 12447101. 
\end{acknowledgments}


\bibliography{apssamp}

\end{document}